 \documentclass[final,3p,times,twocolumn]{elsarticle}

\usepackage{graphicx}
\usepackage{amssymb}
\usepackage{amsmath}
\usepackage{units}
\usepackage{lineno}

\biboptions{sort&compress}

\journal{Nuclear Instruments and Methods in Physics Research A}

\begin{document}
\bibliographystyle{elsarticle-num}

\begin{frontmatter}



\title{A Method to Correct Differential Nonlinearities in Subranging Analog-to-Digital Converters Used for Digital $\gamma$-ray Spectroscopy}


\author[col]{A.~Hennig\corref{cor1}}
\ead{hennig@ikp.uni-koeln.de}
\cortext[cor1]{Corresponding author}
\author[col]{C.~Fransen}
\author[xia]{W.~Hennig}
\author[col]{G.~Pascovici}
\author[col]{N.~Warr}
\author[col]{M.~Weinert}
\author[col]{A.~Zilges}

\address[col]{Institute for Nuclear Physics, University of Cologne, Z\"ulpicher Stra{\ss}e 77, D-50937 Cologne, Germany}
\address[xia]{XIA LLC, 31057 Genstar Rd., Hayward CA 94544, USA}

\begin{abstract}
The influence on $\gamma$-ray spectra of differential nonlinearities (DNL) in subranging, pipelined analog-to-digital converts (ADCs) used for digital $\gamma$-ray spectroscopy was investigated. The influence of the DNL error on the $\gamma$-ray spectra, depending on the input count-rate and the dynamic range has been investigated systematically. It turned out, that the DNL becomes more significant in $\gamma$-ray spectra with larger dynamic range of the spectroscopy system. An event-by-event offline correction algorithm was developed and tested extensively. This correction algorithm works especially well for high dynamic ranges.
\end{abstract}


\end{frontmatter}

\section{Introduction}
\label{sec:introduction}

The HORUS spectrometer (\textbf{H}igh efficiency \textbf{O}bservatory for $\gamma$-\textbf{R}ay \textbf{U}nique \textbf{S}pectroscopy) \cite{Linn05b} is located at the 10 MV Tandem ion accelerator at the Institute for Nuclear Physics in Cologne. It consists of 14 high-purity Germanium (HPGe) detectors for high-resolution $\gamma$-ray spectroscopy. Six of these HPGe detectors can be equipped with BGO shields for active Compton suppression. In addition, two of the standard HPGe detectors can be replaced by  Clover-type detectors \cite{Duch99} for Compton polarimetry experiments \cite{Butl73, Simp83}.

The coincident detection of charged particles in the exit channel of a nuclear reaction and the deexciting $\gamma$-rays provides valuable additional information. For example in inelastic scattering experiments (e.g. (p,p$^{\prime}\gamma$), (d,d$^{\prime}\gamma$)) or transfer reactions (e.g. (p,d$\gamma$)), the excitation energy of the target nucleus can be derived on an event-by-event basis from the energy of the charged particle in the exit channel \cite{Catf86}. For this purpose, the particle detector array SONIC \cite{Pick14} can be embedded within the HORUS spectrometer. SONIC houses up to eight $\Delta$E-E silicon sandwich detectors which allow particle identification.

To process up to 36 detector channels, the analog data acquisition system was replaced by a digital data acquisition system, based on the DGF-4C Rev. F modules manufactured by XIA LLC \cite{Warb00,Skul00,Hubb99}. This is a CAMAC based module that comprises four complete spectroscopic channels. The DGF-4C modules have been extensively used for the data acquisition at the Miniball spectrometer \cite{Warr13}, where Rev. D modules are used. Nowadays, newer revisions of the DGF-4C modules are available, that possess a USB connector for fast data readout. In contrast to the conventional analog signal processing approach, the preamplifier signal is directly digitized and hence, all spectroscopic information, e.g. energy and time information, can be extracted using digital filter algorithms \cite{Jord94}.

With the HORUS spectrometer and the combined setup of HORUS and SONIC manifold aspects of nuclear physics can be studied. As examples we mention two applications: The study of the Pygmy Dipole Resonance (see \cite{Savr13} and references therein) and the in-beam investigation of nuclear reactions relevant for nuclear astrophysics (see \cite{Kaep11, Raus13, Arno07} and references therein).

\begin{figure}[t!]
	\begin{center}
	\includegraphics[width=0.47\textwidth]{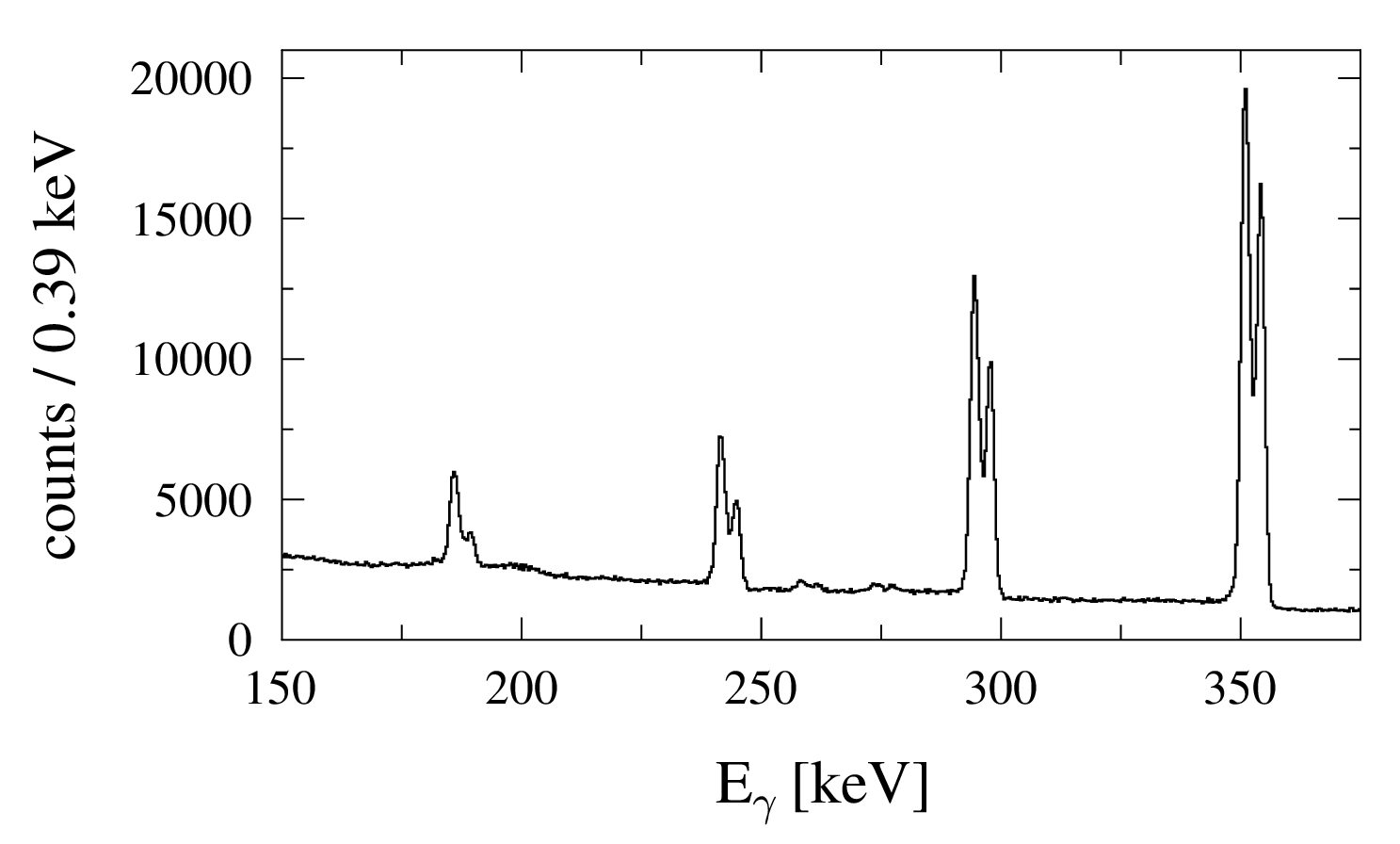}
	\end{center}
\caption{Low-energy part of a $\gamma$-ray spectrum obtained with a $^{226}$Ra calibration source. The spectrum was taken with a 20\% coaxial HPGe detector at an input count-rate of 23 kcps. The dynamic energy range was set to 12.8 MeV. The double-peak structure of the peaks is evident.}
\label{fig:uncorrected}
\end{figure}

Both types of application require the detection of $\gamma$-rays in the range from 5 to 15 MeV. It turned out, that when operating the new data acquisition system at such an energy range (denoted as high dynamic range in the following), significant distortions of the peak shape, e.g. double- or even multiple-peak structures start evolving in the $\gamma$-ray spectra. Fig. \ref{fig:uncorrected} shows an excerpt of a $\gamma$-ray spectrum obtained with a $^{226}$Ra calibration source. The double-peak structure of the peaks is obvious. As already pointed out previously \cite{Vent01, Pasc13, Laue04}, the observed spectral distortions can be traced back to the digitization process, in particular to the differential nonlinearities (DNL) of the analog-to-digital converters (ADC) used in the spectroscopy system.

In this paper, a new method is presented to correct the effect of the DNL of subranging ADCs in the $\gamma$-ray spectra by means of an offline correction-algorithm. The correction algorithm requires a calibration procedure of each individual channel, together with an online pulse-shape analysis (PSA).

In section 2, the principle of digital pulse-processing with the DGF-4C is shortly presented. In section 3 the effect of the DNL on measured $\gamma$-ray spectra will be investigated systematically. The newly developed correction algorithm will be presented in section 4 and the results achieved with the new algorithm will be discussed in section 5.

\section{Signal processing in the DGF-4C}
\label{sec:processing}
In this section, the signal processing in the DGF-4C is shortly sketched. We will constrain ourselves to the parts which are relevant for the effect and correction of the DNL. A more detailed description of the signal-processing technique in the DGF-4C can be found e.g. in Refs. \cite{Warb00,Skul00,Hubb99}. The preamplifier signals from the semiconductor detectors are characterized by a rising edge with a typical length of a few tens of ns followed by an exponential decay with a time constant of usually about 50~$\mu$s. The signal is directly connected to the input channels of the DGF-4C modules. The signal passes an analog gain and offset stage as well as a Nyquist filter, before it is digitized in an 80 MHz 14 bit sampling ADC, for which the ADC AD6645 from the company Analog Devices is used. Since this component is the origin of the DNL it will be further discussed in section 3.

The digital filtering of the digitized preamplifier-signal is based on the moving window deconvolution (MWD) technique \cite{Geor93}, which is implemented on field programmable gate arrays (FPGAs). The height of the preamplifier signal, which contains the information on the energy deposited in the active volume of the detector can be obtained by taking the average of several samples of the digitized signal before and after the rising edge of the signal. This can be expressed by the following equation 

\begin{equation}
\label{eq:trapez}
V=-\sum\limits_{j=k-2L-G+1}^{k-L-G}V_j + \sum\limits_{j=k-L+1}^k V_j,
\end{equation} 

\noindent where $L$ denotes the number of samples to be averaged while $G$ accounts for the width of the rising edge of the input signal. $V_k$ is the ADC value $L$ samples after the rising edge. The value $V$ is continuously calculated from the ADC sample stream $V_j$ \cite{Skul00}. Applying this type of filter to a step-like input-signal results in a trapezoidal-shape filter response \cite{Jord94, Skul00} with the width of the rising and falling slope of the trapezoid $L$ and the width of the flat top $G$. Two trapezoidal-shape filter algorithms are applied to the digitized preamplifier signal: A fast filter, which has a rather short filter length (usually the filter parameters $L$ and $G$ are in the range of a few tens of ns), is used for triggering, time determination and pile-up rejection. The height of the input signal is extracted using a slow filter, for which $L$ and $G$ are typically in the range of $1-10~\mu$s and $0.5-1~\mu$s, respectively. 

If an event is validated at the end of the slow filter, the filter sums are read out by a digital signal processor (DSP), which then determines the energy and time information from the provided filter sums, including a correction for ballistic deficit. Furthermore, the DSP is capable of applying algorithms for an online pulse-shape analysis (PSA), if parts of the digitized preamplifier signals have been captured in the FIFO. Finally, onboard pulse-height spectra are recorded by the DSP and listmode data are written to buffers, which then can be read out via the USB interface. In addition, it is possible to capture and read out a series of digitized samples via a First In First Out buffer (FIFO). 

In order to operate several modules synchronously, the clock of one module can be defined as a master and can be distributed among the others. Global event building for all modules is finally performed with an offline sorting-algorithm.

\section{Systematic investigation of the DNL and its effect on $\gamma$-ray spectra}
\label{sec:analysis}

The effect of the DNL on the peak shape and linearity was investigated using pulses, produced by a BNC PB-5 programmable pulse generator and with a standard coaxial HPGe detector with a relative efficiency of 20 \% from the company ORTEC.

\begin{figure}[t!]
	\begin{center}
	\includegraphics[width=0.47\textwidth]{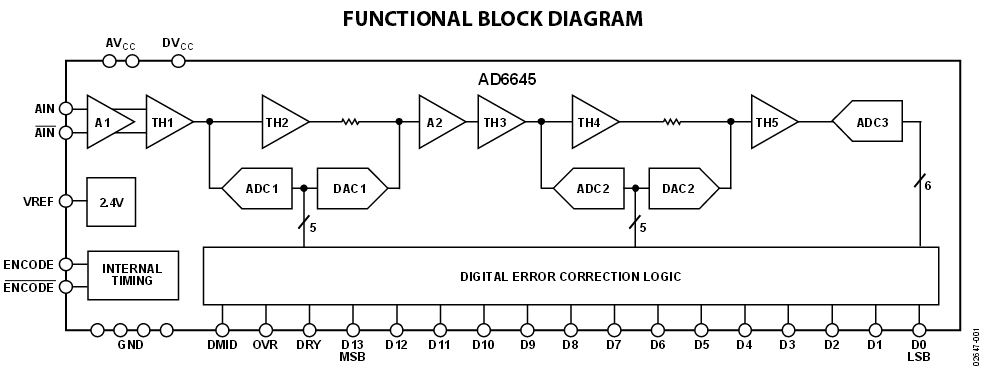}
	\end{center}
\caption{Simplified block diagram of the 14bit, 80MSPS ADC AD6645 which is used for the digitization of the preamplifier signal in the DGF-4C, from \cite{Kest06}. The subranging, pipelined ADC includes two 5-bit ADC's (ADC1 and ADC2) followed a 6-bit ADC (ADC3). Significant DNL distortions occur at the ADC1 transition points \cite{Kest06}.}
\label{fig:block}
\end{figure}

Fig. \ref{fig:block} shows a simplified block diagram of the AD6645 which is used in the DGF-4C modules of Rev. F. The AD6645 is a subranging, pipelined ADC, that includes a 5-bit ADC1, followed by a 5-bit ADC2 and a 6-bit ADC3. Two of the bits are used for a digital error correction logic, so that the effective depth of the whole ADC amounts to 14 bit.

The DNL of this specific ADC was already measured by W. Kester \cite{Kest06} (see especially Fig. 9 in Ref. \cite{Kest06}). There are $2^5=32$ ADC1 transition points which are $2^9=512$ ADC channels apart. Significant DNL errors were found at these transition points with the expected spacing of 512 LSB. In this particular case, the DNL error amounts to about 1.5 LSB. In a schematic way,  the effects on the transfer function of the ADC (analog input code vs. digital output code) in presence of a DNL are pointed out, which should be of step-like character (see Fig. 8 in Ref. \cite{Kest06}).

In \cite{Pasc13} the DNL of the AD6645 was measured in a different manner. A signal from a pulse generator with the shape of a preamplifier signal and of variable height was uniformly distributed over the top half of the ADC range. In case of no DNL, the resulting pulse-height spectrum is a uniform distribution. In contrast, deviations of up to 5\% of the average number of counts per bin have been observed.

\begin{figure}[t!]
	\begin{center}
	\includegraphics[width=0.47\textwidth]{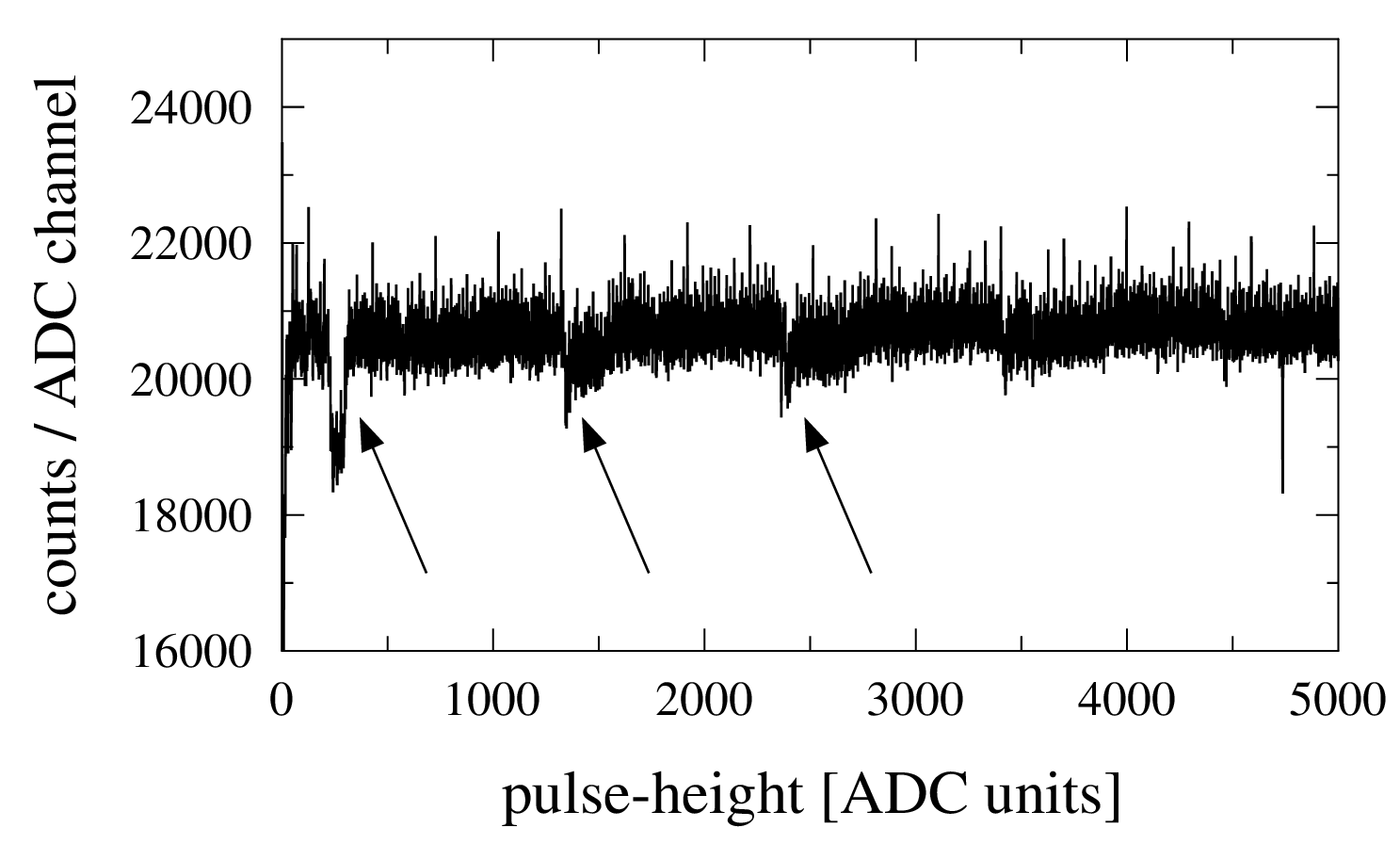}
	\end{center}
\caption{Pulse-height spectrum obtained with a high-precision pulse generator in the range of the lowest 5000 ADC channels. The pulse-height is ramped uniformly across the ADC range. A measurement with an ADC showing no DNL would result in a flat spectrum. In contrast, regular distortions are observed, indicated by the arrows.}
\label{fig:ramped}
\end{figure}

To confirm these results a similar measurement was performed with the pulse generator connected to the DGF-4C, where a preamplifier-like pulse of variable height was uniformly covering the whole ADC range. The resulting pulse-height spectrum is shown in Fig. \ref{fig:ramped}, showing similar results to those obtained in \cite{Pasc13} (see Fig. 6 in \cite{Pasc13}).

Nevertheless, DNL errors tend not to be limited to this particular ADC only. The effect of the DNL using other types of fast pipeline-ADCs has been investigated as well \cite{Vent01}. Multi-peak structures have been observed, when analysing the pulse height of signals of fixed amplitude generated by a pulse generator, starting from different offset values \cite{Vent01}. Hence, the DNL effects become more significant at high count rates, where the baseline varies from one pulse to another as each pulse can be located on the tail of a previous one \cite{Pasc13}.
  
\begin{figure}[t!]
	\begin{center}
	\includegraphics[width=0.47\textwidth]{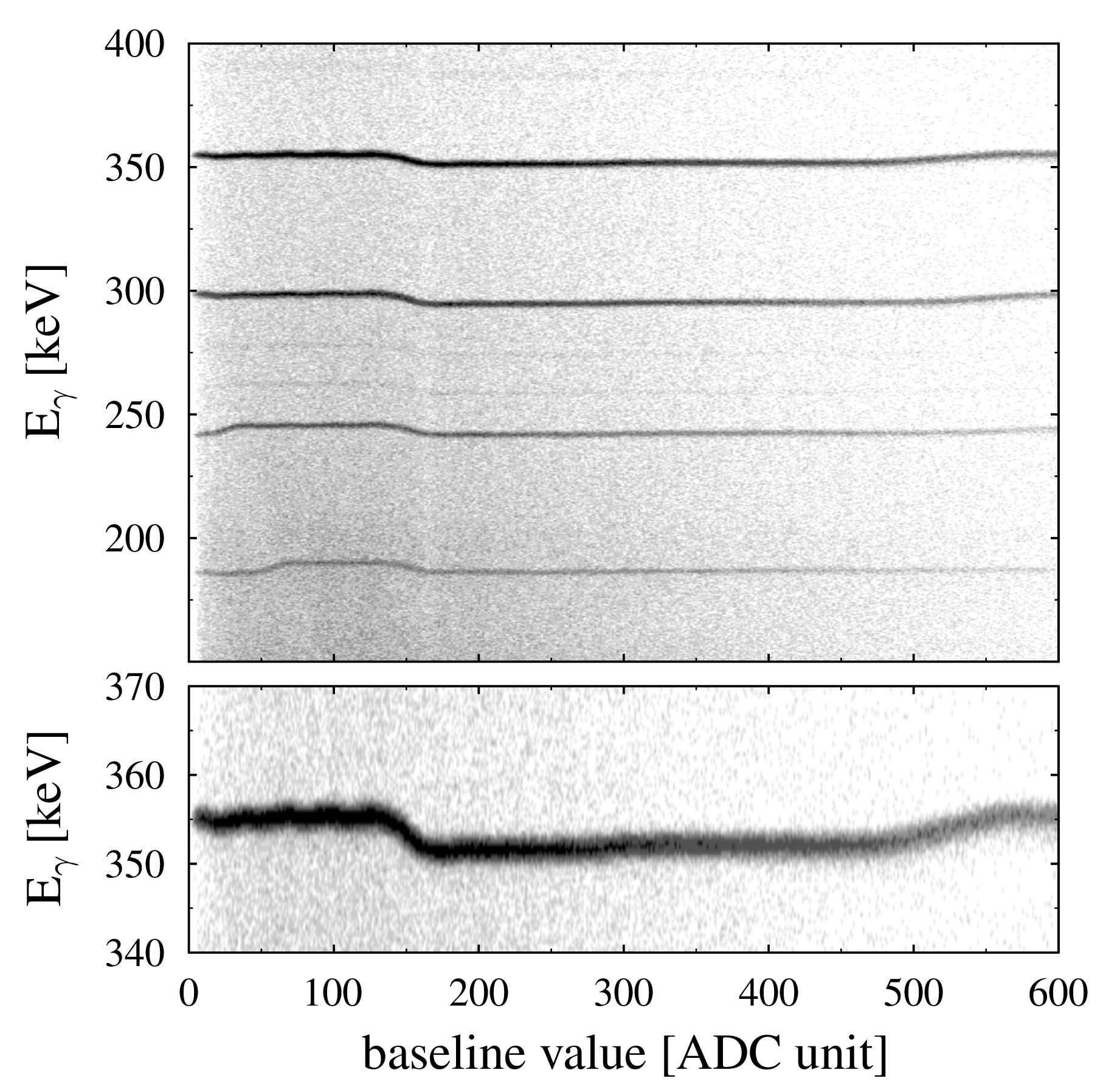}
	\end{center}
\caption{Intensity plot of the measured $\gamma$-ray energy versus the baseline value (see text for further details). A projection of the matrix towards the y-axis would result in a $\gamma$-ray spectrum. The observed double-peak strucutre originates from the discontinuities, showing up at e.g. a baseline value of 168 ADC units. The lower panel is a closer look to the energy-region around 354~keV.}
\label{fig:matrix}
\end{figure}

\begin{figure}[t!]
	\begin{center}
	\includegraphics[width=0.47\textwidth]{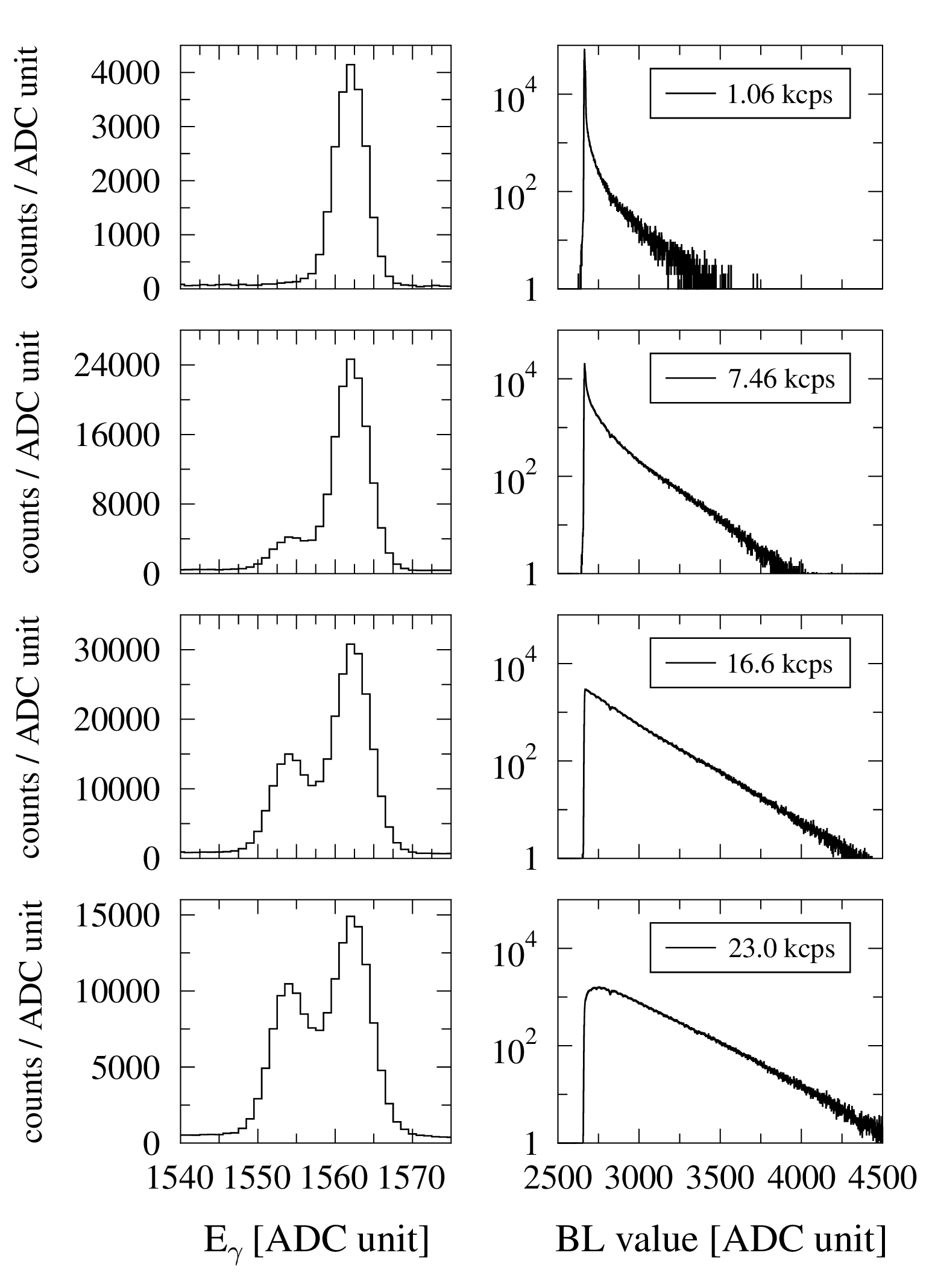}
	\end{center}
\caption{Evolution of the spectral distortions with increasing count rate. The left panel shows the $\gamma$-ray spectrum around the peak at 609 keV, stemming from the  $^{226}$Ra calibration source. The right panel shows the corresponding baseline distribution, which gets broader with increasing input count-rate. The double-peak structure gets more pronounced with increasing width of the baseline distribution.}
\label{fig:countrate}
\end{figure}

To further elaborate the count rate dependency and thus, the dependency on the baseline, waveforms were captured in a measurement with the HPGe detector using a $^{226}$Ra calibration source. From this measurement an average ADC value right before the rising edge of the pulse was extracted to obtain a baseline value for each individual event. In Fig. \ref{fig:matrix}, these baseline values are shown together with the energy values, computed by the DGF-4C in a 2D-matrix. This figure confirms the observations from \cite{Vent01} and \cite{Pasc13}, that the double-peak structures seen  in the $\gamma$-ray spectrum (which would be a projection of the matrix to the vertical axis) arise from baseline variations due to high count rates. Fig. \ref{fig:countrate} shows a systematic investigation of the count-rate dependence. The left panel shows the energy spectrum (y-axis projection of Fig. \ref{fig:matrix}) around the 609-keV $\gamma$-ray transition stemming from the $^{226}$Ra calibration source for different input count-rates together with the corresponding baseline distribution (x-axis projection of Fig. \ref{fig:matrix}) in the right panel. The double-peak structure is more pronounced with increasing width of the baseline distribution.

\begin{figure}[t!]
	\begin{center}
	\includegraphics[width=0.47\textwidth]{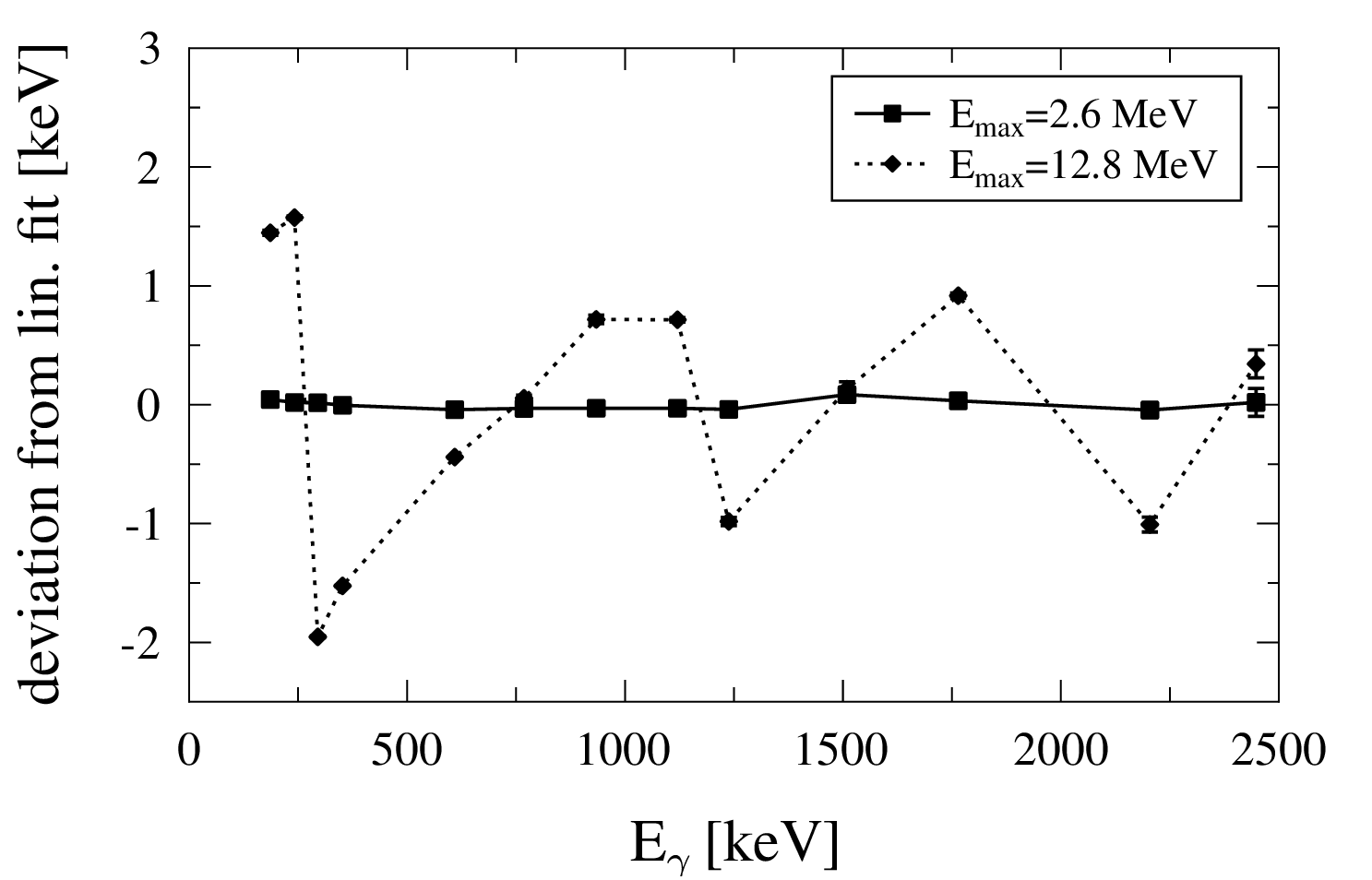}
	\end{center}
\caption{Deviation of the measured peak positions from a linear fit to the data. For a dynamic range of 12.8 MeV (connected with dotted lines) the deviation reaches 2 keV and reflects the error function of the transfer function of the ADC 6645 (Fig. 8 in ref \cite{Kest06}). Almost no deviation is observed in the measurement with a dynamic range of 2.6~MeV (connected with solid lines).}
\label{fig:uncorr_lin}
\end{figure}

The uppermost panel of Fig. \ref{fig:countrate} shows a nearly undistorted $\gamma$-ray peak in case of a narrow baseline distribution (low count rates). Nevertheless, from the discussion in \cite{Kest06}, one would expect that the DNL also causes integral nonlinearity (INL), i.e. deviations from a linear energy calibration. For an input count-rate of 1 kcps, using a $^{226}$Ra calibration source and an energy range of 12.8~MeV, Fig. \ref{fig:uncorr_lin} shows the deviation from a linear energy calibration as a function of $\gamma$-ray energy (dotted lines). Deviations up to $\pm 2~\mathrm{keV}$ are observed. It has to be emphasized, that this deviation occurs, although no double-peaks are observed in the $\gamma$-ray spectra. In contrast, the deviation has almost completely vanished for a dynamic range of 2.6 MeV (solid lines).

The effect of the DNL depending on the number of bits of the sampling ADC has been already discussed \cite{Vent01}. To be more precise, one should discuss the effect more likely in terms of number of ADC values per energy range instead of total number of bits of the sampling ADC only. Simply speaking, an ADC with a depth of 12~bit and a dynamic range of 4~MeV would give the same spectral distortions as a 14-bit ADC operated with a maximum energy range of 16~MeV. Therefore, it is intuitive to introduce the ADC channel width $W$ in absolute energy units of keV, defined as the ratio of dynamic range and the number of possible ADC values: Operating a 14-bit ADC at a dynamic range of 16 MeV would correspond to a channel width of $W=0.98$~keV. Nevertheless, since the ADC used in the measurements of this paper is a 14-bit ADC, only a change of the dynamic energy range will affect the channel width in the following.

\begin{figure}[t!]
	\begin{center}
	\includegraphics[width=0.47\textwidth]{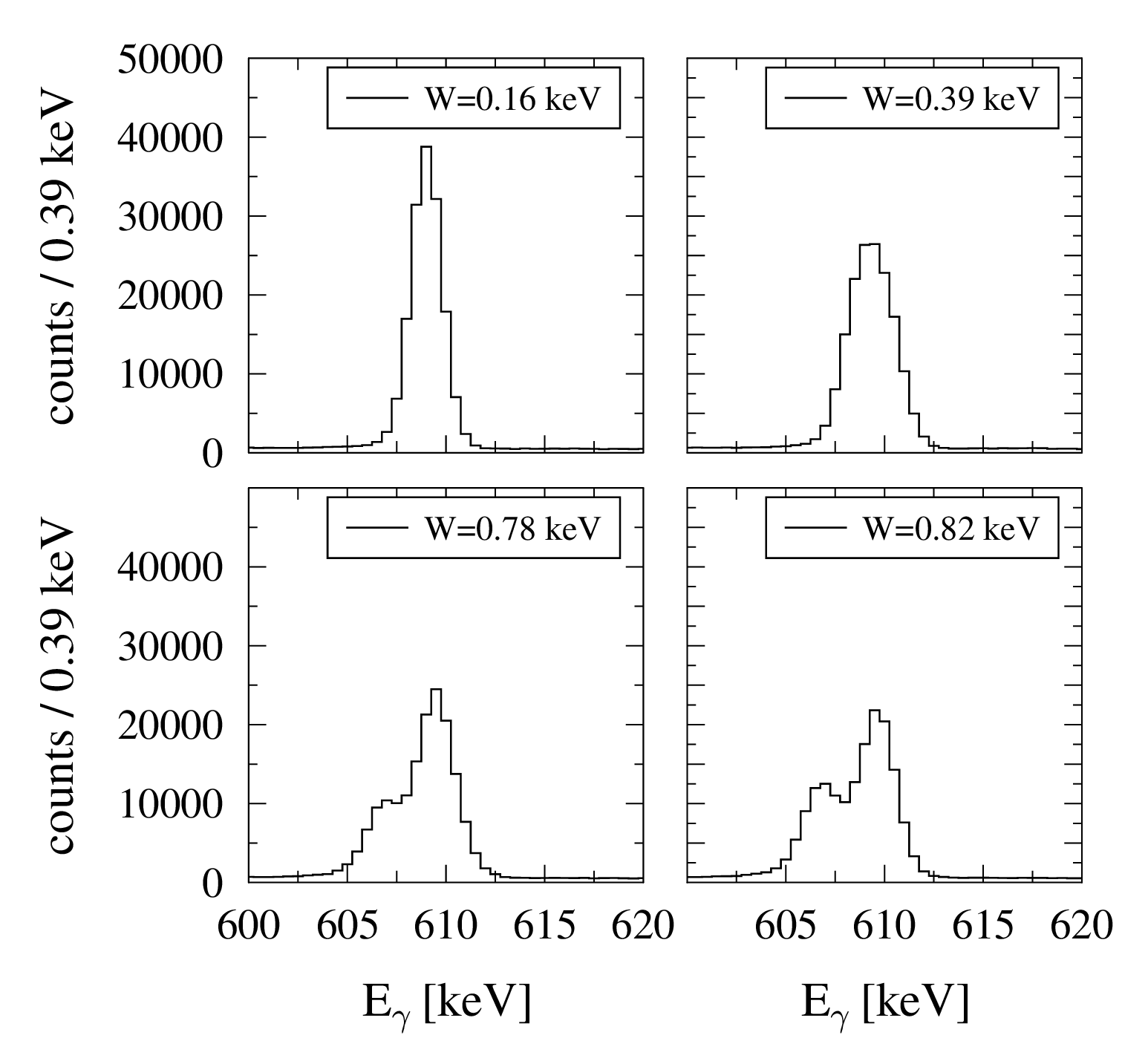}
	\end{center}
\caption{Evolution of spectral distortions with increasing channel width $W$. The individual figures show an excerpt of the spectrum around the 609-keV peak. With increasing channel width, the peak is broadened ($W=0.39~$keV) and finally, double-peak structures start evolving ($W=0.78~$keV and $W=0.82~$keV). The input countrate was 23.9 kcps for all measurements.}
\label{fig:gain}
\end{figure}

The influence of the DNL is strongly reduced with increasing number of ADC values per energy range, i.e. with reduced channel width \cite{Vent01}. This is illustrated in Fig. \ref{fig:gain}, where the energy spectrum in the region of the 609-keV $\gamma$ transition from a $^{226}$Ra calibration source is shown for different channel widths. The degrading of the peak shape is evident.

To conclude the investigations on the DNL, the effect gets more significant with increasing number of ADC-values per energy range. If the baseline distribution is broadened, double-peak structures evolve for large channel widths. Though no double-peak structures appear for narrow baseline distributions, deviations from a linear energy calibration are observed for large channel widths.

\section{The DNL correction algorithm}

A common technique to overcome DNL in ADCs is the sliding-scale method \cite{Cott63} and in the case of fast pipeline-ADCs the dithering technique \cite{Kest06}. In the latter case, an additional noise signal is added to the analog input signal. After the digitizing process, the additional signal is substracted from the digitized sum-signal. The net effect is, that the the DNL errors of the ADC are randomly spread across the ADC range. Despite the impressive results of this technique \cite{Kest06}, it is not applicable in this case, since the amplitude of the additional signal to be added would have to span the whole ADC1 range \cite{Laue04}. Adding such an amount of noise would inacceptably affect the energy resolution of the $\gamma$-spectroscopy system.

In a different approach \cite{Laue04}, the digital value obtained in each subranging ADC is extended by two fractional bits, which are fitted to the $\gamma$-ray spectra and stored in a look-up table. Though this technique yields impressive results for individual cases, this approach turned out to be impractical in the present application: Implementing a look-up table on the presently used FPGA would by far exceed its available memory, whereas a transfer of the samples to the DSP to perform a correction at that level would infer a large amount of deadtime. On the other hand, an offline correction of each ADC sample would require the storage of a huge amount of digitized samples on hard-disk and thus inferring a large amount of deadtime as well. However, this technique might be applicable with the development of FPGAs with larger capacity.

For the correction presented in this work, an off\-line event-by-event correction algorithm was developed. The principle of this algorithm is to compare the position of each individual preamplifier pulse with the regions of the ADC showing DNL errors. The location of the preamplifier pulse in the ADC range is determined using the pulse-shape analysis (PSA) mode of the DGF-4C modules. For every pulse, the PSA mode is used to average a couple of sampling points right before the rising edge of the pulse and to average a couple of sampling points right after the rising edge of the pulse. Both values are computed online for each pulse and stored additionally in the listmode data. The first value will be denoted as \textit{baseline value}, the latter as \textit{peak value} in the following.

The correction procedure can be divided into two parts: In a first step, a calibration of each input channel has to be performed to locate the regions of the ADC where the DNL error is present. In a second step, the position of each pulse within the ADC range is compared to these regions and the pulse height is adequately corrected. The two steps will be discussed in more detail in the next sections.

\subsection{Calibration procedure}

\begin{figure}[t!]
	\begin{center}
	\includegraphics[width=0.47\textwidth]{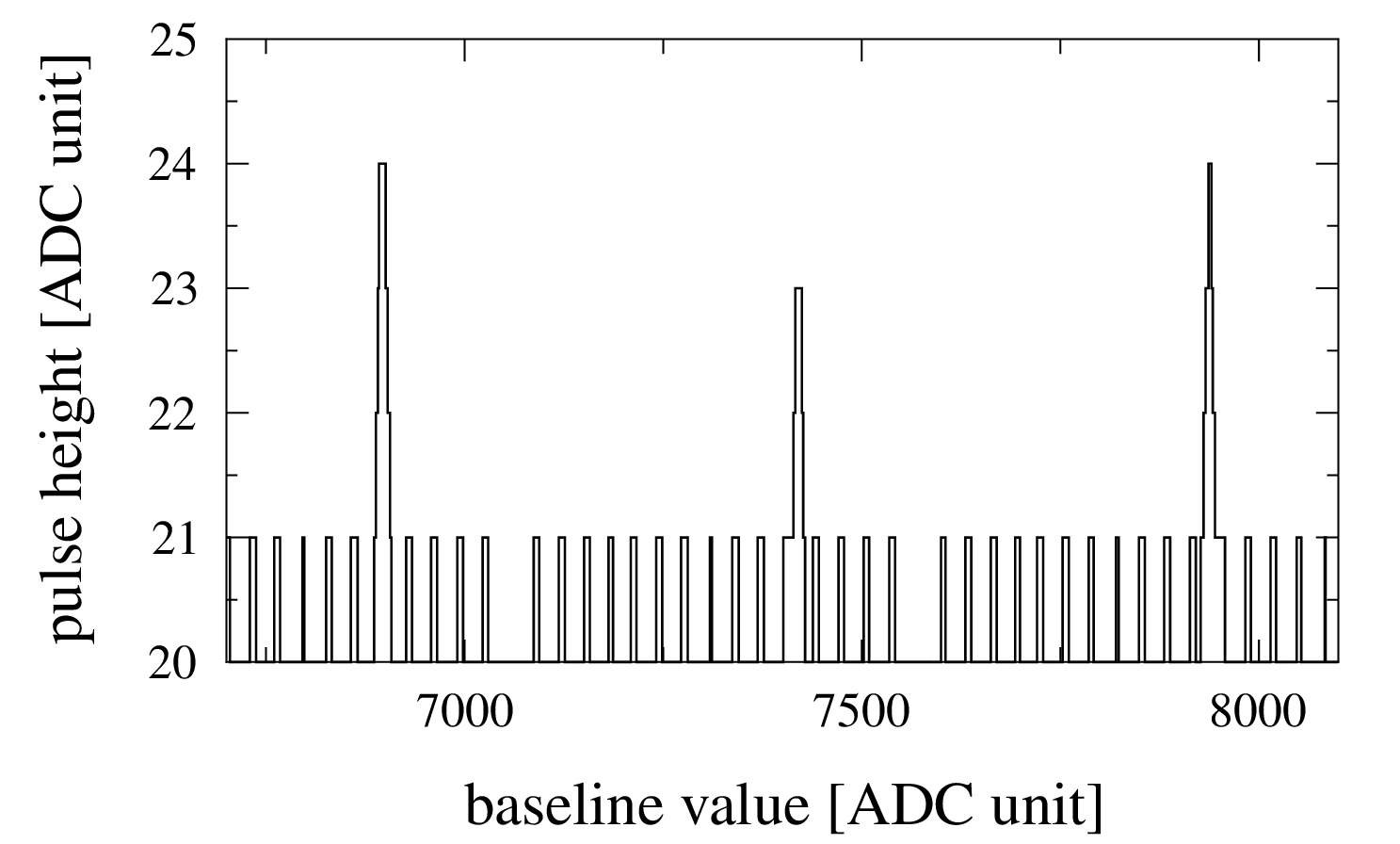}
	\end{center}
\caption{Section of the measured pulse height as a function of the baseline value. A preamplifier-like signal produced with a high-precision pulse-generator was fed into one input channel of a DGF-4C. The offset of the signal was constantly increased using the internal offset stage of the DGF-4C. 32 peaks are observed, corresponding to the 32 ADC1 transition points. Each of the individual peak shows a trapezoidal shape.}
\label{fig:lattenzaun}
\end{figure}

For the calibration procedure, a preamplifier-like signal produced by a pulse generator is fed into the DGF-4C. If the complete signal is in a range of the ADC with no DNL error, a pulse height $A$ will be measured by the DGF-4C. To calibrate the ADCs of each input channel of the DGF-4Cs, the value for the internal offset stage is systematically increased and for each value a pulse-height spectrum is obtained. If the pulse reaches an ADC range with DNL error, a pulse height $A^{\prime}\neq A$ will be measured. In addition to the energy value, the peak and baseline values, as described in the previous section are measured. Fig. \ref{fig:lattenzaun} shows the measured pulse height as a function of the baseline value. Since the individual pulses from the pulser are well separated and of constant height for each offset value, the peak and baseline values are constant. With increasing offset value, the baseline value increases as well. The DNL-error regions are well separated from each other. In total, $N=32$ DNL-error regions are observed, with a spacing of 512 ADC channels. This corresponds to the expected number and spacing of ADC1 transitions (see section \ref{sec:analysis}).

Fig. \ref{fig:lattenzaun} reveals a trapezoidal shape for each of the 32 peaks. The rising slope of this trapezoid arises, if the top of the pulse partly covers the DNL-error region in the ADC. If the pulse completely covers the region a flat top emerges. If the offset is further increased, the bottom of the pulse only partly covers the DNL region. This results in the falling slope of the trapezoid. However, because of the small pulse-height used in this measurement an almost triangular shape is obtained for the peaks in Fig. \ref{fig:lattenzaun}. A similar diagram is obtained for the pulse-height as a function of the peak value. The upper and lower boundaries $A_i^h$ and $A_i^l$ of the $i$-th DNL-error region ($i\in \{0,..,N\})$ can be obtained from the lower left corner of the $i$-th trapezoid in the baseline vs. pulse-height diagram and from the lower right corner in the peak vs. pulse-height diagram, respectively. Furthermore, the height of the trapezoid $C_i$ is the error in pulse height caused by the DNL if the pulse completely covers the DNL-error region.

For each input channel of the data acquisition system, a look-up table is generated containing the values $A_i^h$, $A_i^l$ and $C_i$ for each DNL-error region. This look-up table is then used for the correction of pulse-height spectra. It has to be noted, that for the calibration procedure, the height of the pulse, generated by the pulser has to be large enough, to completely span a DNL-error region, but small enough, to not reach two regions at once. The calibration procedure has to be performed only once for every input channel. Every subsequent measurement can then be corrected with this look-up table and the correction algorithm, described in the next section. 

\subsection{Correction of pulse-height spectra}
Measured $\gamma$-ray spectra are corrected by means of an offline-correction algorithm. For each pulse, the peak and baseline value is compared to the look-up table, obtained from the calibration procedure. Every time the pulse crosses a region affected by the DNL, the pulse height is reduced by the amount by which it deviates from an unaffected region. In the following, the first and last DNL region, which is crossed by the pulse is denoted with the index $f$ and $l$, respectively. Hence, $f$ is the lowest index for which $A_f>b$ and $l$ is the highest index for which $A_l<p$ holds, if $b$ and $p$ are the baseline and peak values obtained from the PSA and $A_i$ is the average position of the $i$-th region affected by the DNL:

\begin{equation}
\label{eq:cent}
A_i=\frac{A_i^l+A_i^h}{2}.
\end{equation}

\noindent Thus, the corrected energy value is given as

\begin{equation}
\label{eq:corr}
E^{\mathrm{corr}}=E^{\mathrm{raw}}-\sum\limits_{i=f}^{l}C_i 
\end{equation}

\noindent where $E^{\mathrm{raw}}$ is the non-corrected energy value and $C_i$ is the difference in pulse height between the affected and unaffected ADC regions. Equation (\ref{eq:cent}) approximates the regions affected by the DNL as a single ADC value, although they are observed to span a range of three to four ADC values. Nevertheless, this approximation is validated by the effects of finite filter-lengths used for the trapezoidal-filter algorithm, which will be discussed in the following.

\begin{figure}[t!]
	\begin{center}
	\includegraphics[width=0.47\textwidth]{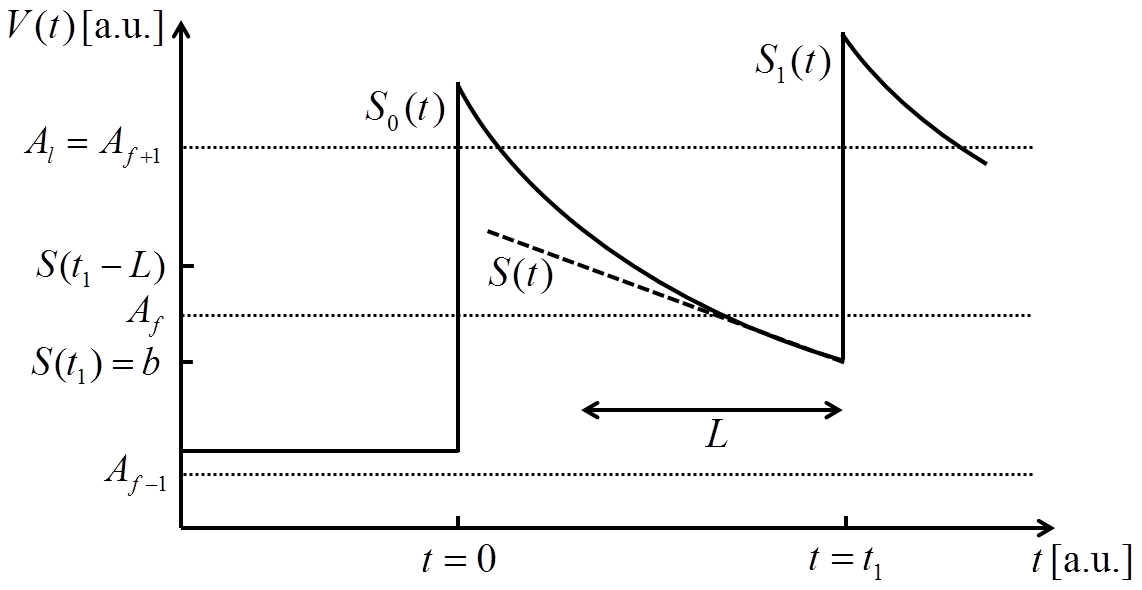}
	\end{center}
\caption{Two schematic preamplifier signals $S_0(t)$ and $S_1(t)$ (solid line) as a function of time. $S_0(t)$ starts at $t=0$ while $S_1(t)$ is located on the trailing edge of $S_0(t)$. The sum signal is linearly approximated in the vicinity of $t=t_1$. Because of the finite filter-length $L$, sampling-points enter into the sums of equation \ref{eq:trapez}, which are above and below the first relevant DNL region at $A_f$. Note that the x axis is not to scale.}
\label{fig:summing}
\end{figure}

The deduced pulse height of the input-signal is obtained from the trapezoidal-filter algorithm (see equation (\ref{eq:trapez})) presented in section \ref{sec:processing}. The $i$-th region of the DNL can affect the pulse-height determination, if $V_n\leq A_i\leq V_m$ for every $n$ and $m$ which are indices entering in the first or second sum of equation (\ref{eq:trapez}), respectively. In this case, the energy value can be corrected by means of equation (\ref{eq:corr}). This is no longer sufficient, if the sums of equation (\ref{eq:trapez}) contain values that are both, larger and smaller than $A_i$, i.e. if

\begin{equation}
\label{eq:ocrbl}
V_{k-2L-G+1} \geq A_i \geq V_{k-L-G}
\end{equation}

\noindent along with $b<A_i$, or

\begin{equation}
\label{eq:ocrpeak}
V_{k-L+1} \geq A_i \geq V_{k}
\end{equation}

\noindent along with $p>A_i$. Equation (\ref{eq:ocrbl}) corresponds to those cases, where the pulse is located on the trailing edge of a previous pulse, so that the first filter sum of equation (\ref{eq:trapez}) contains ADC values above and below the discontinuity (see Fig. \ref{fig:summing}). This is most likely to occur for the DNL region at $A_f$. Similarly, equation (\ref{eq:ocrpeak}) corresponds to the case, where the pulse only slightly exceeds the discontinuity, but due to the exponential decay of the pulse height, the second filter sum of equation (\ref{eq:trapez}) contains again ADC values located above and below the discontinuity. This is most likely to occur for the DNL region at $A_l$. In both cases, the energy value would be over-corrected so that equation (\ref{eq:corr}) has to be modified to

\begin{equation}
\label{eq:ocrcorr}
E^{\mathrm{corr}}=E^{\mathrm{raw}}-\sum\limits_{i=f}^{l}C_i + w_fC_f + w_lC_l
\end{equation}

\noindent including weighting factors $w_f$ and $w_l$, that depend on how many sampling points contained in the filter-sums are above and below the discontinuity.

In order to determine the weighting factor $w_f$, assume two pulses $S_0(t)$ and $S_1(t)$, starting at $t=0$ and $t=t_1$, with amplitudes $N_0$ and $N_1$ and a common time constant $\tau$ (see Fig. \ref{fig:summing}):

\begin{equation}
S_0(t)=N_0\exp\left(-\frac{t}{\tau}\right)
\end{equation}

\begin{equation}
S_1(t)=N_1\exp\left(-\frac{t-t_1}{\tau}\right)
\end{equation}

\noindent $N_0$ is unknown, while $N_1$ can be written as $N_1=p-b$. For $L\ll\tau$, $S_0(t)$ and $S_1(t)$ can be approximated using a first order Taylor approximation in point $t_1$ (see Fig. \ref{fig:summing}), so that the summed signal can be written as

\begin{equation}
S(t)=S_0(t)+S_1(t)\approx b\left(1-\frac{t}{\tau}-\ln\left(\frac{b}{N_0}\right)\right)
\end{equation}

\noindent Since $S(t)$ is linear in $t$ within this approximation, $w_f$ can be written as

\begin{equation}
w_f=\frac{S(t_1-L)-A_0}{S(t_1-L)-S(t_1)},
\end{equation}

\noindent and finally, using $S(t_1)=b$:

\begin{equation}
\label{eq:wf}
   w_f =
   \begin{cases}
     0 & \mathrm{if }~ A_f > b\cdot\left(1+\frac{L}{\tau}\right) \\
     1+ \frac{b-A_f}{b\frac{L}{\tau}} & \mathrm{if }~ A_f \leq b\cdot\left(1+\frac{L}{\tau}\right)
   \end{cases}
\end{equation}

\noindent Since $A_f>b$, $w_f$ is smaller than one. In a similar way, an expression for $w_l$ can be obtained:

\begin{equation}
\label{eq:wl}
   w_l =
   \begin{cases}
     0 & \mathrm{if }~ A_l < p\cdot\left(1-\frac{L}{\tau}\right) \\
     1+ \frac{A_l-p}{p\frac{L}{\tau}} & \mathrm{if }~ A_l \geq p\cdot\left(1-\frac{L}{\tau}\right)
   \end{cases}
\end{equation}

Equations (\ref{eq:wf}) and (\ref{eq:wl}) yield the important result, that $w_f$ and $w_l$ are independent of the unknown height of the previous pulse $N_0$. It has to be emphasized, that this is no longer true, if $S_0(t)$ and $S_1(t)$ are approximated taking into account quadratic or even higher-order terms of $t/\tau$.

\section{Results}

\begin{figure}[t!]
	\begin{center}
	\includegraphics[width=0.47\textwidth]{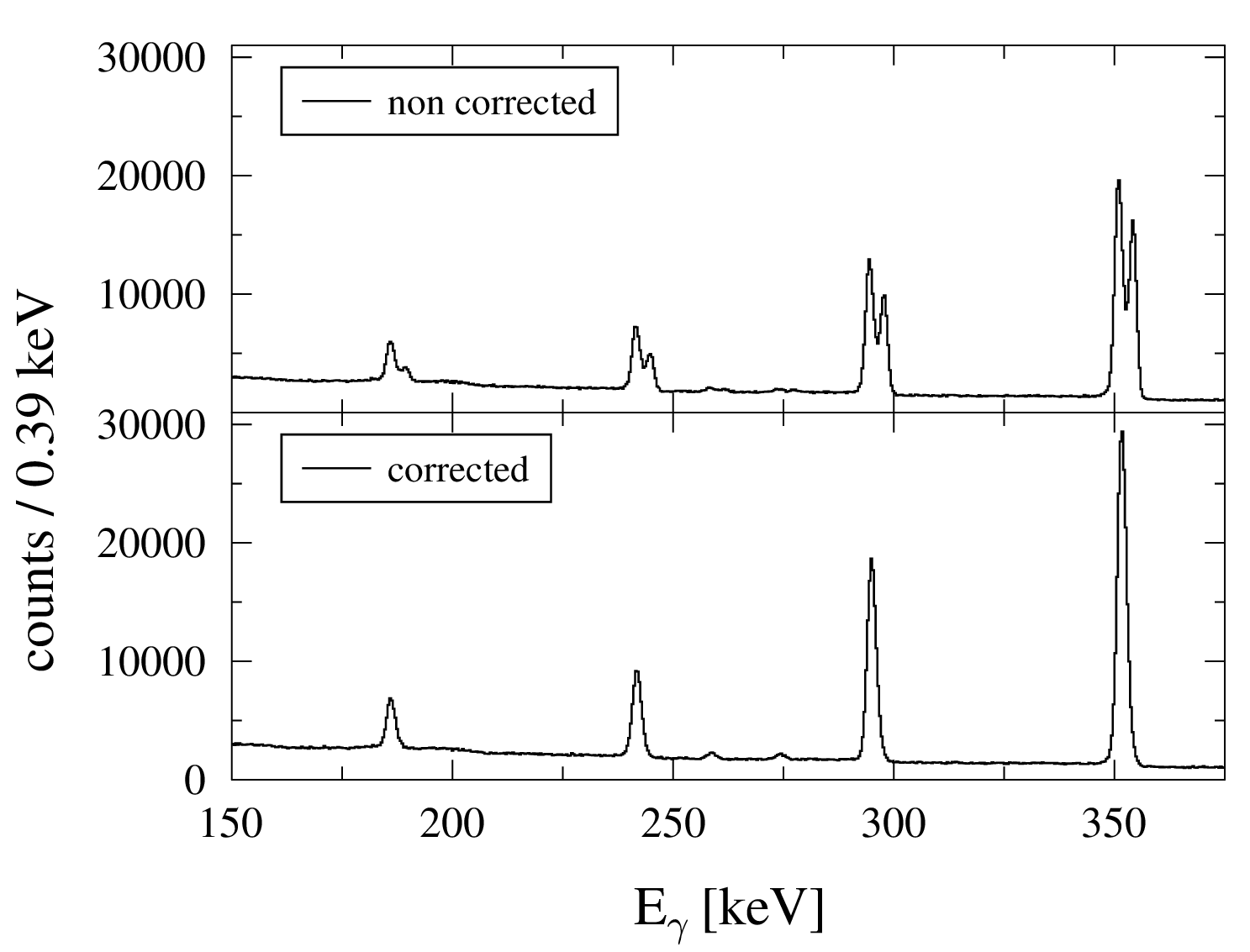}
	\end{center}
\caption{Low-energy part of the raw $\gamma$-ray spectrum (upper panel) and the corrected $\gamma$-ray spectrum (lower panel). The measurement was performed at an input count-rate of 23~kcps and a channel width of 0.78~keV. Using the correction algorithm, the Gaussian peak-shape can be restored. Note that both y axes have the same scale.}
\label{fig:226Ra_corr}
\end{figure}

The correction algorithm has been tested using a standard coaxial HPGe detector from the company ORTEC with a relative efficiency of 20 \% along with a standard $^{226}$Ra calibration source which provides $\gamma$-rays in the energy range from 180~keV up to 2447~keV. Fig. \ref{fig:226Ra_corr} shows a section of the low-energy part of a raw $\gamma$-ray spectrum (upper panel) together with the spectrum which is obtained, by applying the correction algorithm (lower pannel). The spectrum was taken at a count rate of 23 kcps and a dynamic range of 12.8 MeV, corresponding to a channel width of $W=0.78$~keV. Using the correction algorithm, the double-peak structure can be removed.

\begin{figure}[t!]
	\begin{center}
	\includegraphics[width=0.47\textwidth]{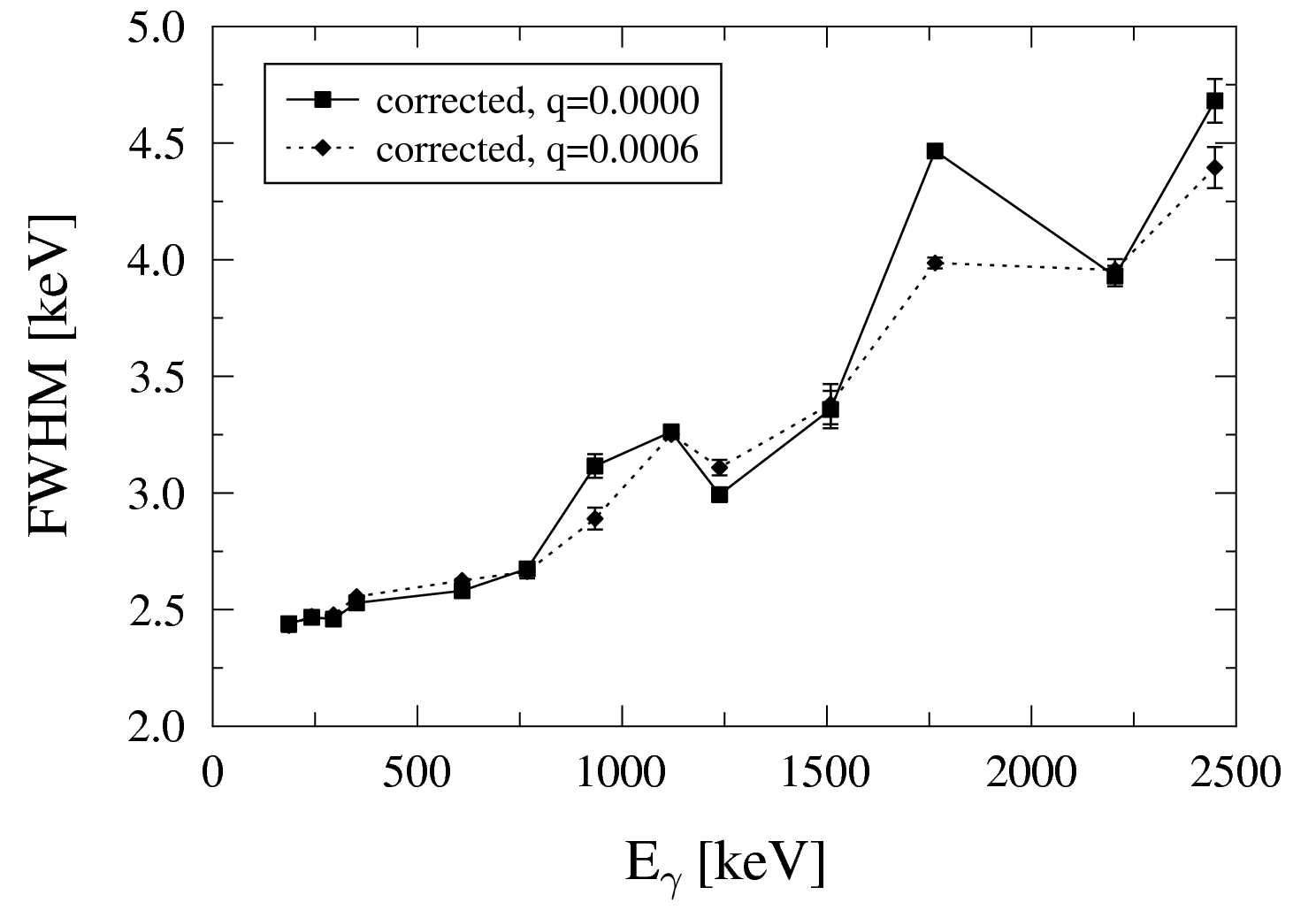}
	\end{center}
\caption{Measured energy resolution as a function of $\gamma$-ray energy. The data points connected with solid lines are obtained using the correction algorithm without taking a quenching factor into account, thus using $q=0$. The data-points connected with the dotted lines are obtained with a value of $q=0.0006$. The energy-resolution for the peaks at 964~keV, 1764~keV and 2447~keV can be significantly improved by introducing the quenching factor. The data were taken with a channel width of 0.78~keV and an input count-rate of 23~kcps. Note that no comparison to non-corrected values can be drawn for this case because of the observed double-peak structure.}
\label{fig:fwhm_rate1}
\end{figure}

\begin{figure}[t!]
	\begin{center}
	\includegraphics[width=0.47\textwidth]{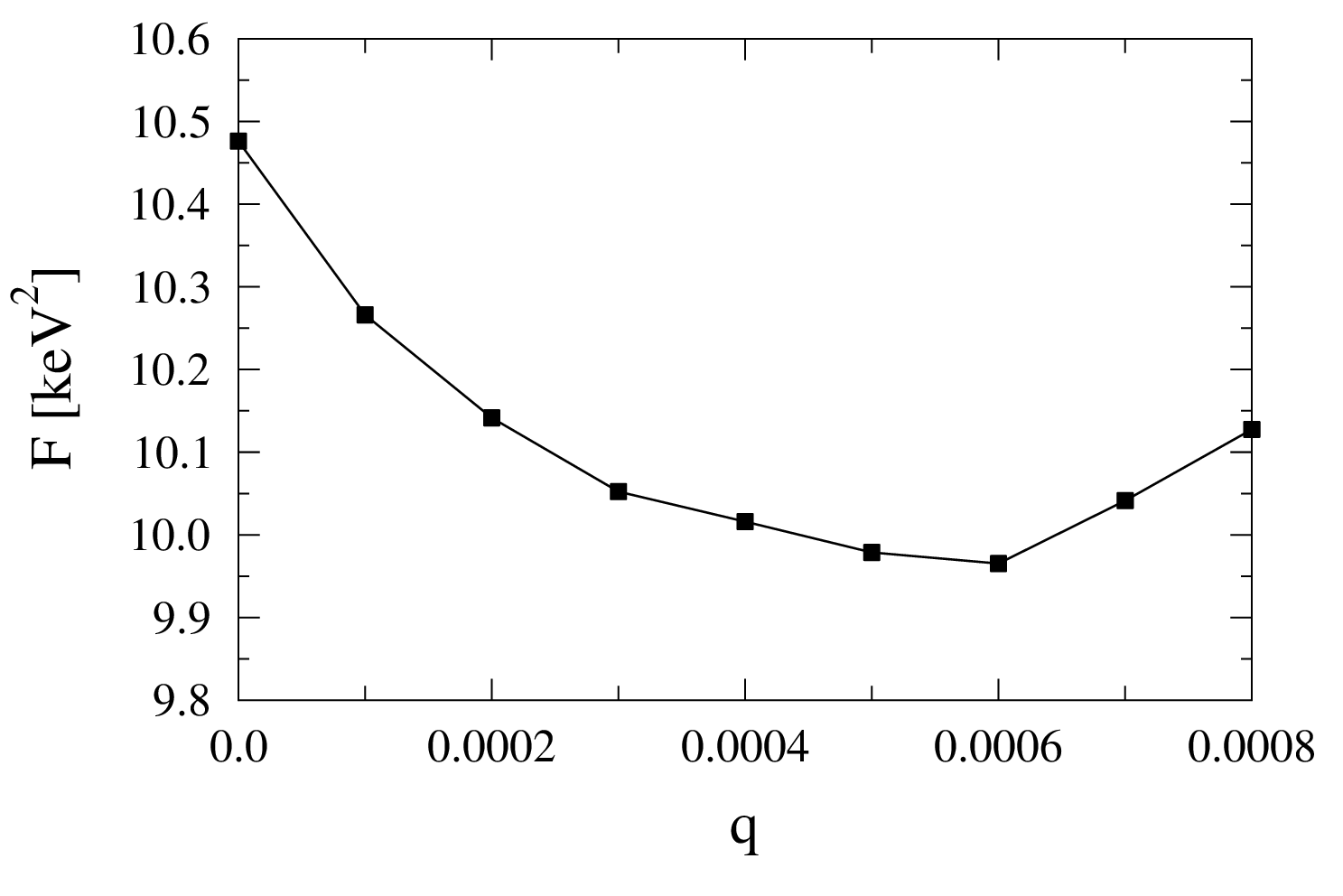}
	\end{center}
\caption{Fit of the quenching factor $q$ to minimize the figure of merit $F$, defined in equation \ref{eq:fom}. $F$ is minimized for a value of $q=0.0006$.}
\label{fig:fom_fwhm}
\end{figure}

\begin{figure}[t!]
	\begin{center}
	\includegraphics[width=0.47\textwidth]{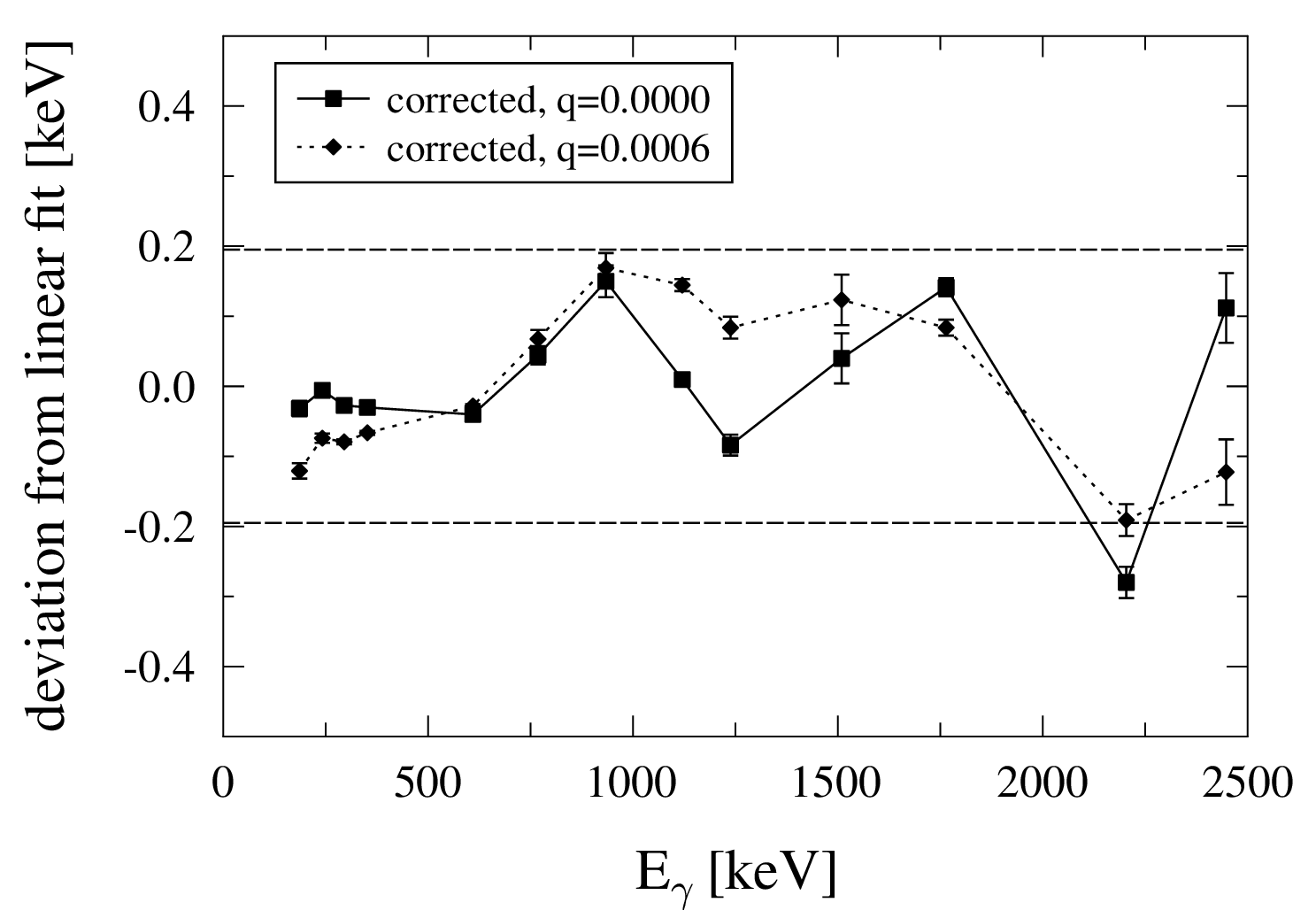}
	\end{center}
\caption{Deviation of the corrected peak positions from a linear fit to the data using a quenching factor $q=0$ (solid line) and $q=0.0006$ (dotted line). The data were taken at a count rate of 23~kcps and a channel width of 0.78~keV. The two horizontal lines indicate the width of one energy bin of the $\gamma$-ray spectrum. The introduction of the quenching factor has only minor influence on the linearity. Nevertheless, the deviation of the peak position is found to be smaller than one energy bin.}
\label{fig:lin_rate1}
\end{figure}

The algorithm described in the previous section assumes a few simplifications, i.e. the exponential decay of the preamplifier signal is linearly approximated and the finite rise-time is neglected. Furthermore, in the trapezoidal filter (see equation (\ref{eq:trapez})) implemented in the DGF-4C also contributions from the rising edge are taken into account for ballistic deficit corrections. Fig. \ref{fig:fwhm_rate1} shows the peak-width (FWHM) of several peaks in the corrected $\gamma$-ray spectrum as a function of energy. It can be recognized, that at certain energies, the peak is broadened compared to the neighbouring ones (e.g. at $E_{\gamma}=1764~$keV and $E_{\gamma}=2447~$keV). This may result from the simplifications sketched above. To improve the results, an additional quenching factor $q$ was introduced and the weighting factor $w_l$ in equation (\ref{eq:wl}) is modified in the case of $A_l \geq p\cdot\left(1-\frac{L}{\tau}\right)$ to 

\begin{equation}
   w_l = 1+ \frac{A_l-p}{p\frac{L}{\tau}}+\frac{1}{C_f}q\cdot E^{\mathrm{raw}} 
\end{equation}

\begin{figure}[t!]
	\begin{center}
	\includegraphics[width=0.47\textwidth]{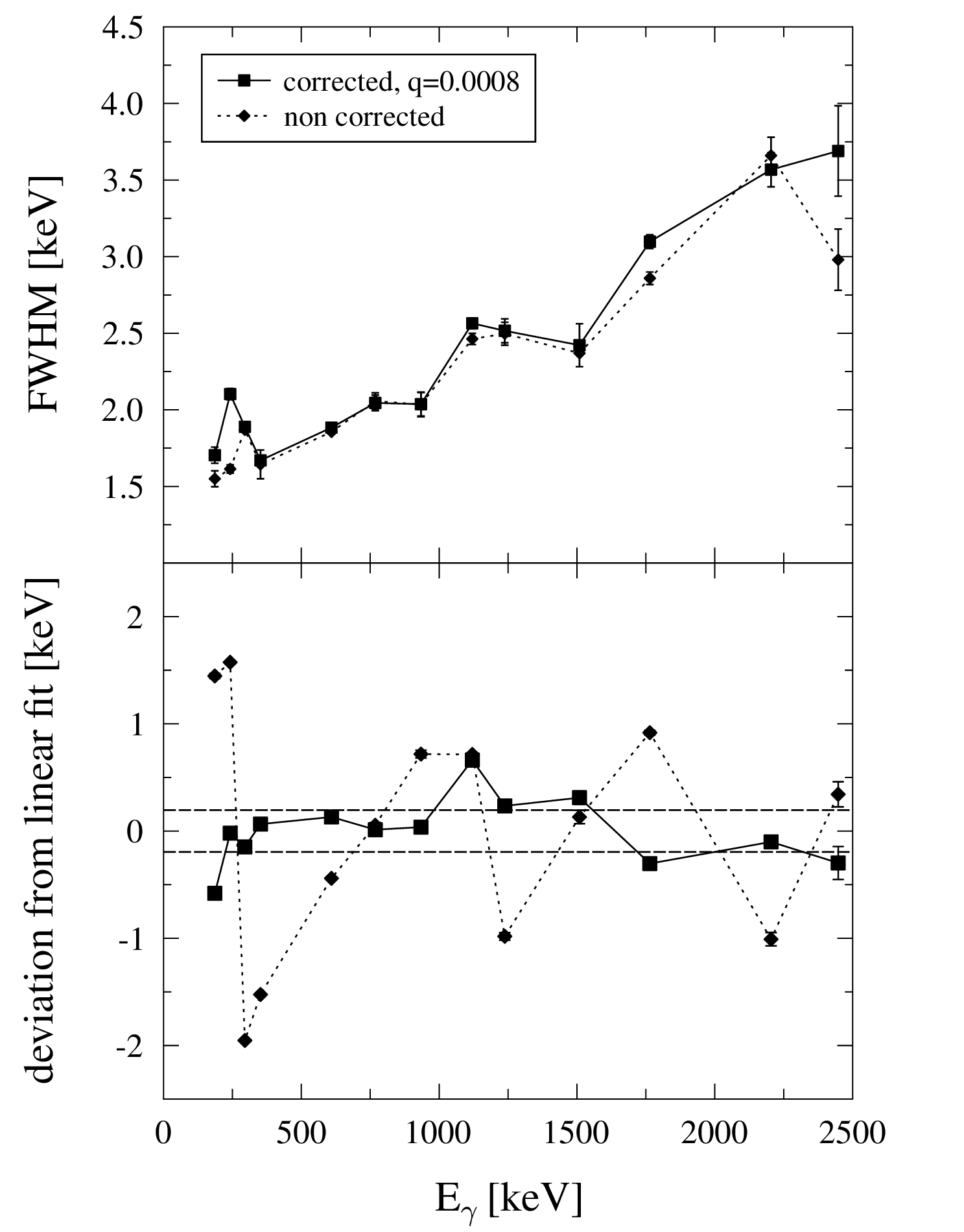}
	\end{center}
\caption{Same as Fig. \ref{fig:fwhm_rate1} (upper panel) and \ref{fig:lin_rate1} (lower panel), for an input count-rate of 1.06 kcps and compared to the non-corrected results. In terms of energy resolution, the results for the non-corrected and the corrected spectra are nearly the same, while the linearity is significantly improved using the correction algorithm.}
\label{fig:res_caseII}
\end{figure}

\begin{figure}[t!]
	\begin{center}
	\includegraphics[width=0.47\textwidth]{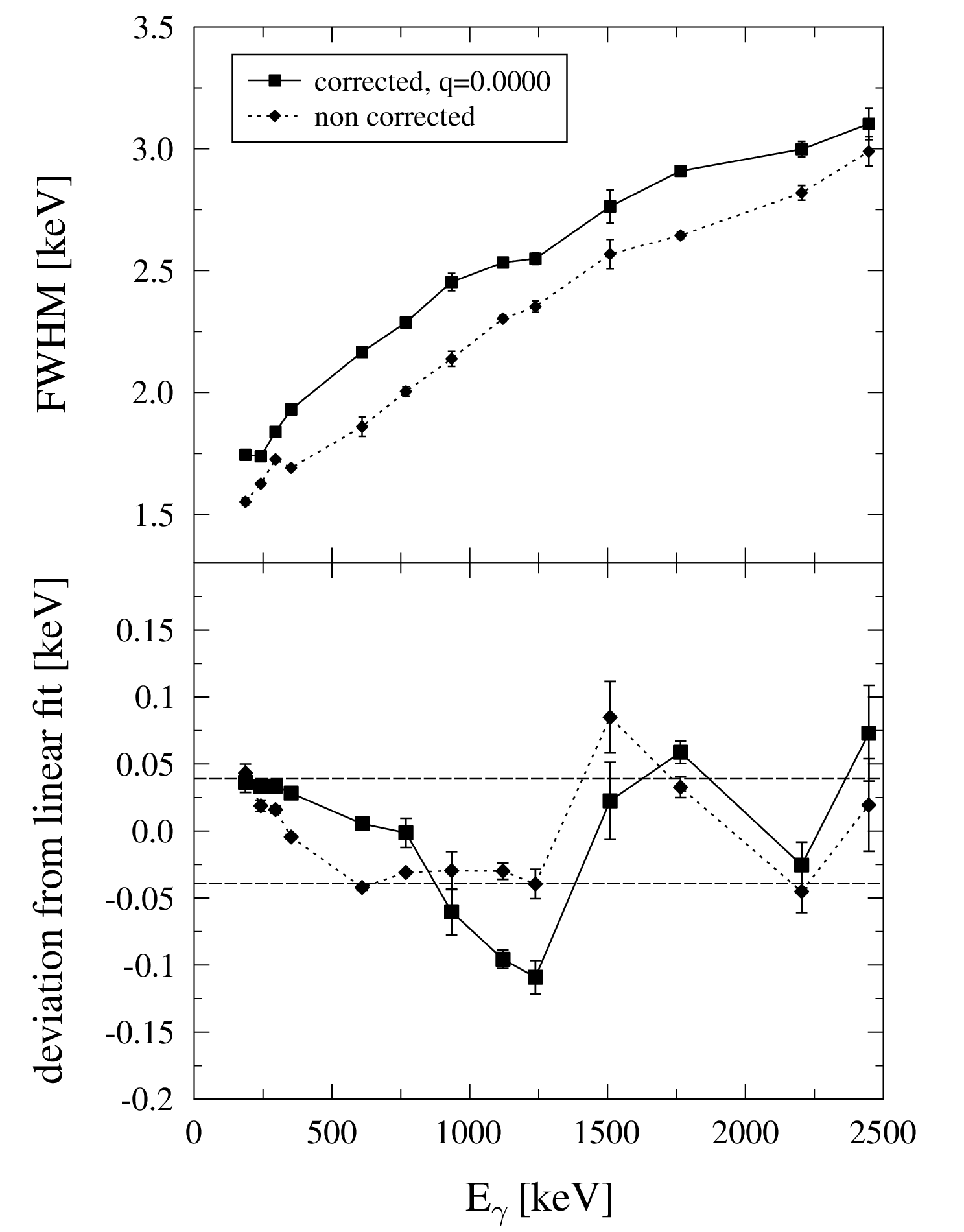}
	\end{center}
\caption{Same as Fig. \ref{fig:fwhm_rate1} (upper panel) and \ref{fig:lin_rate1} (lower panel), for a channel width of 0.16~keV and compared to the non-corrected results. The energy resolution is worsened using the correction algorithm by 0.2 - 0.3 keV in the whole energy range. In terms of linearity, the corrected and non-corrected spectra yield the same results.}
\label{fig:fwhm_lin_others}
\end{figure}

\noindent i.e. the correction is additionally reduced linearly with energy. $q$ can be fitted to obtain the best overall energy-resolution. For this purpose, a figure of merit $F$ has been defined as

\begin{equation}
\label{eq:fom}
F=\frac{1}{Z}\sum\limits_{i=1}^{Z}\sigma_i^2
\end{equation}

\noindent where $\sigma_i$ are the peak-widths (FWHM) of a total number of $Z$ peaks in the $\gamma$-ray spectrum. Hence, the optimum value for $q$ is obtained for a minimum value of $F(q)$. Fig. \ref{fig:fom_fwhm} shows $F$ as a function of $q$, yielding a minimum at $q=0.0006$. Using this value for the correction algorithm, the peak width of the outliers can be significantly reduced, while the widths of the other peaks remain almost unchanged (see Fig. \ref{fig:fwhm_rate1}). It has to be emphasized that because of the double-peak structures, no comparison of the peak width to the non-corrected values can be made.

Besides the energy resolution, the corrected spectra were investigated with respect to the integral linearity. Fig. \ref{fig:lin_rate1} shows the deviation of their peak position from a linear fit. The dashed horizontal lines indicate the width of one bin in the energy spectrum, indicating that the deviation is smaller than this. The quenching factor $q$ has only a minor effect on the integral linearity.

The measurement described above was performed with a channel width of 0.78~keV in combination with a count rate of 23~kcps. Reducing the count rate to a value of 1.06~kcps while keeping the dynamic range constant leads to a narrower baseline distribution and the double-peak structure vanishes. Hence, the corrected and non-corrected spectra can be compared in terms of peak width and linearity. The lower panel of Fig. \ref{fig:res_caseII} reveals significant deviations from a linear fit for the non-corrected peak positions. Though the two peaks at 244~keV and 298~keV are only about 50~keV apart, they deviate by about 2~keV from the linear fit in opposite directions. The linearity is significantly improved by the correction algorithm. When comparing the peak-width in the corrected and non-corrected spectra (Fig. \ref{fig:res_caseII} upper panel), no major differences can be observed. The quenching factor has only a minor effect on the peak widths in this case.

Finally, a spectrum has been taken for a low channel width of 0.16~keV, corresponding to a dynamic range of 2.6~MeV, and a high count rate of 24 kcps. As noted in section \ref{sec:analysis}, the effect of the DNL should be less significant in this case. The results in terms of linearity and peak width are shown in Fig. \ref{fig:fwhm_lin_others}. It turns out that the linearity is almost unaffected by the correction algorithm, leading to a maximum deviation from a linear fit, that slightly exceeds the width of one energy bin for the corrected and non-corrected spectra. In contrast, the correction algorithm worsens the peak width by an almost constant value of about $0.3~$keV.

\section{Conclusions}
A correction algorithm for differential nonlinearities in subranging, pipelined analog-to-digital converters used for digital $\gamma$-ray spectroscopy has been developed and tested. The algorithm is especially successful in restoring Gaussian peak-shapes in the case of measurements with large dynamic ranges, i.e. large channel widths. Additionally, the integral linearity can be significantly improved in the case of low count rates. Nevertheless, for small channel widths, the obtained results in terms of peak width are still slightly worse compared to the non-corrected values.

A special feature of this method in contrast to other approaches \cite{Laue04} is, that the calibration procedure for the look-up table has to be performed only once and can be used for every subsequent experiment. Furthermore, the algorithm incorporates only one parameter to be fitted. These advantages make this method feasible for large- and medium scale experiments with a high number of HPGe detectors. It has to emphasized that the method presented in this work is limited to this particular, but for this application widely used type of ADC. Other ADCs like e.g. Flash-ADCs or Wilkinson-type ADCs need different methods (see e.g. \cite{Cott63}).

\section*{Acknowledgements}
We thank V. Derya and J. Eberth for fruitful discussions. We furthermore thank S. G. Pickstone for carefully reading the manuscript. This work is supported by the Deutsche Forschungsgemeinschaft under Contract ZI 510/4-2 and the Bonn-Cologne Graduate School of Physics and Astronomy.

\section*{References}

\bibliography{bibtex/bibtex.bib}


\end{document}